\documentclass[12pt]{article}

\usepackage{amsmath}
\usepackage{amssymb}
\usepackage{graphicx}
\usepackage[dvips]{color}
\usepackage{colordvi}
\usepackage{url}

\begin{document}

\def\d{\displaystyle}
\def\be{\begin{equation}}
\def\ee{\end{equation}}
\def\bea{\begin{eqnarray}}
\def\eea{\end{eqnarray}}
\def\gsim{\:\raisebox{-0.75ex}{$\stackrel{\textstyle>}{\sim}$}\:}
\def\lsim{\:\raisebox{-0.75ex}{$\stackrel{\textstyle<}{\sim}$}\:}

\begin{center}
{\large\bf GAUGE/GRAVITY DUALITY AND SOME APPLICATIONS}

\vspace{3ex}

SPENTA R. WADIA$^*$

\vspace{3ex}

Tata Institute of Fundamental Research, \\
Homi Bhabha Road, Colaba, Mumbai 400 005, India \\
Conference in Honour of \\
Murray Gell-Mann's 80th Birthday \\
25 February 2010 \\
Nanyang Technological University, Singapore \\
$^*$E-mail: wadia@theory.tifr.res.in

\end{center}

\vspace{3ex}

We discuss the AdS/CFT correspondence in which space-time emerges from
an interacting theory of D-branes and open strings. These ideas have a
historical continuity with QCD which is an interacting theory of
quarks and gluons. In particular we review the classic case of D3
branes and the non-conformal D1 brane system. We outline by some
illustrative examples the calculations that are enabled in a strongly
coupled gauge theory by correspondence with dynamical horizons in
semi-classical gravity in one higher dimension.  We also discuss
implications of the gauge fluid/gravity correspondence for the
information paradox of black hole physics.



\section{Introduction:}\label{aba:sec0}

A look at the history of elementary particles reveals that a majority
of the constituents of the Standard Model, were conceived by theory
before they were experimentally established. This includes quarks that
were predicted by Murray Gell-Mann\cite{one} and George Zweig\cite{oneb}
to explain hadron spectra.  In this note we want to discuss the
elementary constituents of space-time and their interactions from
which emerges a theory of gravitation.  

We hope that the theoretical
conceptions we dwell upon in this note contribute to a quantum theory
of gravity that enables a description of phenomena like the formation
and evaporation process of a black hole and basic issues in cosmology
like the origin and fate of the universe.  Besides being a framework
to address these difficult questions, `string theoretic' methods seem to
have applications to various problems in gauge theories, fluid
dynamics and condensed matter physics.  This note is written with the
hope that the ideas and methods of string theory are accessible to a
large community of physicists.

In order to pose the question better we first consider the more
familiar setting of quantum chromo-dynamics (QCD) where the quarks
interact via color gluons. This theory accounts for the spectrum of
hadrons, their interactions and properties of nuclei. In particular in
the limit of long wave lengths the chiral non-linear sigma
model\cite{two} (another invention of Gell-Mann with Levy) describes
the interactions of pions. This theory is characterized by a
dimensional coupling, the pion coupling constant $f^{-1}_{\pi}$.  In 3+1
dimensions $f_{\pi}$ has the dimensions of $({\rm mass})^2$ and if we
generalize the QCD gauge group from $SU(3)$ to $SU(N)$, then
$f_{\pi}\sim N$. Hence in the limit of large N the pions are weakly
coupled and the theory has soliton solutions with `baryon number'. The
mass of the baryon is proportional to N, the number of quark
constituents\footnote{This theory, which emerges from QCD, had
  phenomenological antecedents in the theory of super-conductivity in
  the work of Nambu and Jona-Lasinio\cite{three,four}.}. In modern
terminology one would say that the chiral model is `emergent' from an
underlying theory of more elementary constituents and their
interactions. The phenomenon of `emergence' occurs in complex systems
in many areas of science, and also social sciences.  Gell-Mann has
also contributed to this area\cite{five}.

{\it Over the last 25 years one question that has occupied theoretical
  physicists is: In what sense is gravity an emergent phenomenon? What
  are the fundamental constituents of `space-time' and their
  interactions}. The question is quite akin to that of the emergence
of the chiral model from QCD. If gravity is the analogue of the chiral
model, where the gravitational (Newton) coupling is dimensional and
akin to the `pion coupling constant', what is the QCD analogue for
gravity, from which gravity is emergent?

Perturbative string theory gave the first hint that gravity may be
derived from a more microscopic theory because its spectrum contains a
massless spin 2 excitation\cite{six,seven}.  However real progress
towards answering this question came about with the discovery of
D-branes\cite{eight}.

\section{D-branes the building blocks of string theory}

A D-p brane (in the simplest geometrical
configuration) is a domain wall of dimension $p$, where $0
\leq p \leq 9$.  It is  
characterized by a charge and it couples to a $(p+1)$ form
abelian gauge field $A^{(p+1)}$, e.g. a D0 brane couples to a
1-form gauge field $A^{(1)}_\mu$, a D1 brane couples to a 2-form
gauge field $A^{(2)}_{\mu\nu}$ etc.  The D-p brane has a brane
tension $T_p$ which is its mass per unit volume.  The crucial point is
that $T_p \propto 1/g_s$.  This dependence on the coupling
constant (instead of $g^{-2}_s$) is peculiar to string theory.  It has
a very important consequence.  A quick estimate of the gravitational
field of a D-p brane gives, $G_N^{(10)} T_p \sim g^2_s/g_s \sim 
g_s$.  Hence as $g_s \rightarrow 0$, the gravitational field goes to
zero! If we stack $N$ D-p branes on top of each other then the
gravitational field of the stack $\sim Ng_s$.  A useful limit to study
is to hold $g_s N = \lambda$ fixed, as $g_s \rightarrow 0$ and $N
\rightarrow \infty$.  In this limit when $\lambda \gg 1$ the stack of
branes can source a solution of supergravity. On the other hand when
$\lambda \ll 1$ there is a better 
description of the stack of $D$-branes in terms of open strings.  
A stack of $D$-branes interacts by the exchange of open strings very
much like quarks which interact by the exchange of gluons.  Fig. 1a 
illustrates the self-interaction of a D2-brane by the emission and
absorption of an open string and Fig. 1b illustrates
the interaction of 2 D2-branes by the exchange of an open string.
In the infra-red limit only the lowest mode of the open
string contributes and hence the stack of $N$ $D$-branes can be
equivalently described as a familiar $SU(N)$ non-abelian gauge theory
in $p+1$ dim.
\begin{center}
\includegraphics[scale=.4]{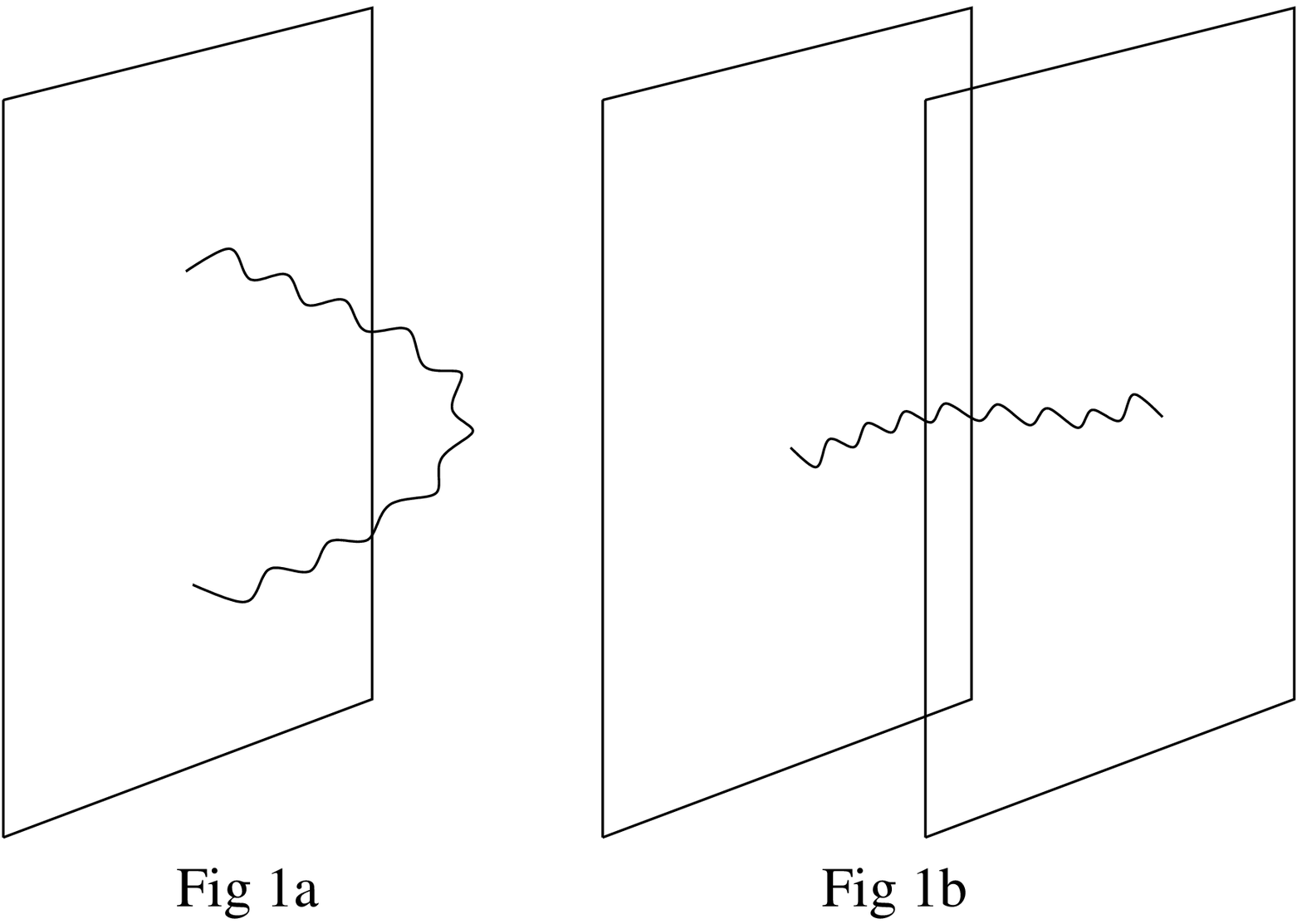}
\end{center}

\section{Statistical Mechanics of D-brane Systems and Black Hole 
Thermodynamics}\label{aba:sec2}

One of the earliest applications of the idea that $D$-branes are the
basic building blocks of `string theory' (and hence of a theory of
gravity) was to account for the entropy and dynamics of certain near
extremal black holes in $4+1$ dim.  As is well known, Strominger and
Vafa\cite{nine} in a landmark paper showed that the
Benkenstein-Hawking entropy of these black holes is equal to the
Boltzmann entropy calculated from the micro-states of a system of D1
and D5 branes
\[
S_{BH} = {A_h \over 4G_N} = k_B \ell n \Omega = S_{\rm Boltzmann}
\]
This result established that black hole entropy can be obtained from
the statistical mechanics of the collective states of the brane system
and it provided a macroscopic basis of the first law of thermodynamics,
$dS_{BH} = T dM$, where $M$ is the mass of the black hole.  Hawking
radiation can be accounted for from the averaged scattering amplitude of
external particles and the micro-states\cite{ten}.

\section{$D3$ branes and the AdS/CFT 
corres\-pondence {\small \cite{eleven,twelve,thirteen,fourteen,fifteen}}:}\label{aba:sec3}

We now discuss the dynamics of a large number $N$, of D3 branes.  A D3
brane is a 3+1 dim. object.  A stack of $N$ D3 branes interacts by the
exchange of open strings.  In the long wavelength limit ($\ell_s
\rightarrow 0$, $\ell_s$ is the string length), only the massless
modes of the open string are relevant.  These correspond to 4 gauge
fields $A_\mu$, 6 scalar fields $\phi^I \ (I = 1,\cdots,6)$
(corresponding to the fact that the brane extends in 6 transverse
dimensions) and their supersymmetric partners.  These massless degrees
of freedom are described by ${\cal N} = 4, \ SU(N)$ Yang-Mills theory
in 3+1 dim.  This is a maximally supersymmetric, conformally invariant
superconformal field theory in 3+1 dimensions.  The coupling constant
of this gauge theory $g_{YM}$, is simply related to the string
coupling $g_s = g^2_{YM}$.  The 'tHooft coupling is $\lambda = g_s N$
and the theory admits a systematic expansion in $1/N$, for fixed
$\lambda$.  Further as $\ell_s \rightarrow 0$ the coupling of the D3
branes to gravitons also vanishes, and hence we are left with the
${\cal N}=4$ SYM theory and free gravitons.

On the other hand when $\lambda \gg 1$, various operators of the large
$N$ gauge theory source a supergravity fluctuations in 10-dimension e.g. the
energy-momentum tensor $T_{\mu\nu}$ sources the gravitational field in
one higher dimension.  The supergravity fields include the metric, two
scalars, two 2-form potentials, and a 4-form potential whose field
strength $F_5$ is self-dual and proportional to the volume form of
$S^5$.  The fact that there are $N$ D3 branes is expressed as
$\d{\int_{S^5}} F_5 = N$.  There are also fermionic fields required by
supersymmetry.  It is instructive to write down the supergravity
metric: \bea ds^2 &=& H^{-1/2} (-dt^2 + d\vec x \cdot d\vec x) +
H^{1/2} (dr^2 + r^2 d\Omega^2_5) \nonumber \\[2mm] && \\ H &=& \left(1
+ {R^4 \over r^4}\right), \ \left({R \over \ell_s}\right)^4 = 4\pi g_s
N \nonumber
\label{fourteen}
\eea
Since $|g_{00}| = H^{-1/2}$ the energy depends on the 5th coordinate
$r$.  In fact the energy at $r$ is related to the energy at $r =
\infty$ (where $g_{00} = 1$) by $E_\infty = \sqrt{|g_{00}|} E_r$.  As
$r \rightarrow 0$ (the near horizon limit), $E_\infty = {r \over R}
E_r$ and this says that $E_\infty$ is red-shifted as $r \rightarrow
0$.  We can allow for an arbitrary excitation energy in string units
(i.e. arbitary $E_r\ell_s$) as $r \rightarrow 0$ and $\ell_s
\rightarrow 0$, by holding a mass scale `$U$' fixed:  
\be
{E_\infty \over \ell_s E_r} \sim {r \over \ell^2_s} = U
\label{fifteen}
\ee 
Note that in this limit the gravitons in the asymptotically flat
region also decouple from the near horizon region.  This is the famous
near horizon limit of Maldacena\cite{twelve} and in this limit the metric
(\ref{fourteen}) becomes \be ds^2 = \ell^2_s \left[{U^2 \over
    \sqrt{4\pi\lambda}} \left(-dt^2 + d\vec x \cdot d\vec x\right) +
  4\sqrt{4\pi\lambda} {dU^2 \over U^2} + \sqrt{4\pi\lambda}
  d\Omega^2_5\right]
\label{sixteen}
\ee
This is locally the metric of ${\rm AdS}_5 \times {\rm S}^5$.  ${\rm AdS}_5$
is the anti-de Sitter space in 5 dim.  This space has a boundary at $U
\rightarrow \infty$, which is conformally equivalent to 3+1
dim. Minkowski space-time.
\bigskip

\noindent{\bf The AdS/CFT conjecture} 

{\it The conjecture of Maldacena is that ${\cal N} = 4$, $SU(N)$ super
Yang-Mills theory 
in 3+1 dim. $\!\!$ is dual to type IIB string theory with ${\rm AdS}_5
\times S^5$ boundary conditions}.  

The gauge/gravity parameters are related as $g^2_{YM}=g_s$ and
$R/\ell_s = (4\pi g^2_{YM} N)^{1/4}$.  It is natural to consider the
$SU(N)$ gauge theory living on the boundary of ${\rm AdS}_5$.  The
gauge theory is conformally invariant and its global exact symmetry
$SO(2,4) \times SO(6)$, is also an isometry of ${\rm AdS}_5 \times
{\rm S}^5$.  The metric (\ref{sixteen}) has a ``horizon'' at $U=0$
where $g_{tt} = 0$.  It admits an extension to the full AdS$_5$
geometry which has a globally defined time like killing vector.  The
boundary of this space is conformal to $S^3 \times {I\!\!\!R}^1$ and
the gauge theory on the boudary is well-defined in the IR since $S_3$
is compact.

The AdS/CFT conjecture is difficult to test because at $\lambda \ll 1$
the gauge theory is perturbatively calculable but the dual string
theory is defined in ${\rm AdS}_5 \times S^5$ with $R \ll \ell_s$.  On
the other hand for $\lambda \gg 1$, the gauge theory is strongly
coupled and hard to calculate.  In this regime $R \gg \ell_s$ and the
string theory can be approximated by supergravity in a derivative
expansion in $\ell_s/R$.  It turns out that for large $N$ and large
$\lambda$, $D$-branes source supergravity fields $\leq$ spin 2.  The
gravitational coupling is given by
\[
G_N \sim g^2_s \sim {\lambda^2 \over N^2} \ll 1
\]
Note the analogy with the constituent formula $f^{-1}_\pi \sim {1 \over N}$. 
The region $\lambda \sim 1$ is most intractable as we can study
neither the gauge theory nor the string theory in a reliable way.
{\it However since the conjecture can be verified for supersymmetric states
on both sides of the duality, one assumes that the duality is true in
general and then uses it to derive interesting consequences for both
the gauge theory and the dual string theory (which includes quantum gravity).}

\noindent {\bf Interpretation of the radial direction of AdS}: 

Before we discuss the duality further we would like to explain the
significance of the extra dimension `$r$'.  Let us recast the ${\rm
AdS}_5$ metric by a redefinition: $\d{r \over R} = e^{-\phi}$ 
\be
ds^2 = e^{-2\phi} \left(-dt^2 + d\bar x \cdot d\bar x\right) +
R^2(d\phi)^2 + R^2 d\Omega^2_5
\label{seventeen}
\ee 
The boundary in these coordinates is situated at $\phi = -\infty$.
Now this metric has a scaling symmetry. For $\alpha > 0$, $\phi
\rightarrow \phi + \log \alpha$, $t \rightarrow \alpha t$ and $\vec x
\rightarrow \alpha \vec x$, leaves the metric invariant.  From this it
is clear that the additional dimension `$Re^\phi$' represents a length
scale in the boundary space-time: $\phi \rightarrow -\infty$
corresponds to $\alpha \rightarrow 0$ which represents 
a localization or short distances in the boundary
coordinates $(\vec x,t)$, while $\phi \rightarrow +\infty$ represents
long distances on the boundary.  $\phi$ is reminiscent of the
Liouville or conformal mode of non-critical string theory, where the
idea of the emergence of a space-time dim. from string theory was
first seen\cite{sixteen}.

The AdS/CFT correspondence clearly indicates that gravity is an
emergent phenomenon.  What this means is that all gravitational
phenomena can be calculated in terms of the correlators of the
energy-momentum tensor of the gauge theory, whose microscopic
constituents are D-branes interacting via open strings.

\section{Black holes and AdS/CFT}

The ${\cal N} = 4$, super Yang-Mills theory defined on $S^3 \times
R^1$ can be considered at finite temperature if we work with euclidean
time and compactify it to be a circle of radius $\beta = 1/T$, where
$T$ is the temperature of the gauge theory.  We have to 
supply boundary conditions which are periodic for bosonic fields and
are anti-periodic for fermions.  These boundary conditions break the
${\cal N} = 4$ supersymmetry, and the conformal symmetry.  However the
AdS/CFT conjecture continues to hold and we will discuss the
relationship of the thermal gauge theory with the physics of black
holes in AdS.

As we have mentioned, in the limit of large $N$ (i.e. $G_N \ll 1$) and
large $\lambda$ (i.e. $R \gg \ell_s$), the string theory is well
approximated by supergravity, and we can imagine considering the
Euclidean string theory partition function as a path integral over all
metrics which are asymptotic to AdS$_5$ space-time.  (For the moment we
ignore $S^5$).

The saddle points are given by the solutions to Einstein's equations
in 5-dim. with a negative cosmological constant
\be
R_{ij} + {4 \over R^2} g_{ij} = 0
\label{twentyfive}
\ee
As was found by Hawking and Page, a long time ago, there are only two
spherically symmetric metrics which satisfy these equations with
AdS$_5$ boundary conditions: AdS$_5$ itself and a black hole solution.

It was shown in Ref. [15] that the `deconfinement' phase of the
gauge theory corresponds to the presence of a large black hole in AdS.
The temperature of the black hole is the temperature of the
deconfinement phase.  The AdS/CFT correspondence says that the
equilibrium thermal properties of the gauge theory in the regime when
$\lambda \rightarrow \infty$ are the same as those of the black
hole. This correspondence enables us to make precise quantitative
statements about the gauge theory at strong coupling $(\lambda \gg
1)$, using the fact that on the AdS side the calculation in gravity is
semi-classical.

We list a few exact results of thermodynamics of the gauge theory at
strong coupling\cite{fourteen}.
\begin{enumerate}
\item[{(i)}] the temperature at which the first order
confinement-deconfinement transition occurs:
\be
T_c = {3 \over 2\pi R_{S^3}}
\label{six}
\ee
where $R_{S^3}$ is the radius of $S^3$.
\item[{(ii)}] the free energy for $T > T_c$
\[
F(T) = - N^2 {\pi^2 \over 8} T^4
\]
\end{enumerate}

Here we see a typical use of the AdS/CFT correspondence calculations
in the strongly coupled gauge theory $(\lambda \gg 1)$ which can be done
using the correspondence by using semi-classical gravity since $G_N
\sim {1 \over N^2} \ll 1$ and ${R \over \ell_s} \sim \lambda^{1/4} \gg
1$.


\vspace{3ex}

\noindent {\bf Conformal Fluid Dynamics and Dynamical Horizons}

We have seen that the thermodynamics of the strongly coupled gauge
theory in the limit of large $N$ and large $\lambda$ is calculable, in
the AdS/CFT correspondence, using the thermodynamic properties of a
large black hole in AdS$_5$ with horizon $r_h \gg R$.  Similar results
hold for a black brane, except that in this case the gauge theory in
$R^3 \times S^1$ is always in the deconfinement phase since $T_c = 0$
if $R_{S^3} = \infty$, by Eq. (\ref{six}).  We now discuss how this
correspondence can be generalized to real time dynamics in this gauge
theory when both $N$ and $\lambda$ are large.

Let us generalize black brane (hole) thermodynamics to fluid
dynamics.  In conformal fluid dynamics the system is in local thermodynamic
equilibrium over a length scale $L$ so that $L \gg {1 \over T}$.  In
the bulk theory in one higher dim. this corresponds to a horizon that is
a slowly varying (see Fig. 2) function of the boundary co-ordinates
$(\vec x,t)$.
\[
r_h \rightarrow r_h + \delta r_h(\vec x,t)
\]
\[
T \rightarrow T + \delta T(\vec x,t)
\]
\[
{1 \over T} {\partial \over \partial x^\mu} {\delta T \over T} \sim {1 \over LT} \ll 1
\]
The ripples on the horizon of a black brane at the linerized level are
analysed in terms of quasi-normal modes with a complex frequencies
$\omega = \omega_R + i\omega_I, \ \omega_I \propto T$, where $T$ is
the temperature of the non-fluctuating brane.  The complex frequency
arises because of the presence of a horizon when we impose only
`in-falling' boundary conditions.  For the dual gauge theory the
quasi-normal mode spectrum implies the dissipation of a small
disturbance of the fluid in a characteristic time.  This is the
qualitative reasoning behind the calculation of `transport
coefficients' of the gauge theory like viscosity, thermal and heat
conductivity which can be done using semi-classical gravity and the
Kubo formula for retarded Green's functions of the corresponding
conserved currents.  This important step was taken by Policastro, Son
and Starinets\cite{seventeen}.

\begin{center}
\includegraphics[scale=.4]{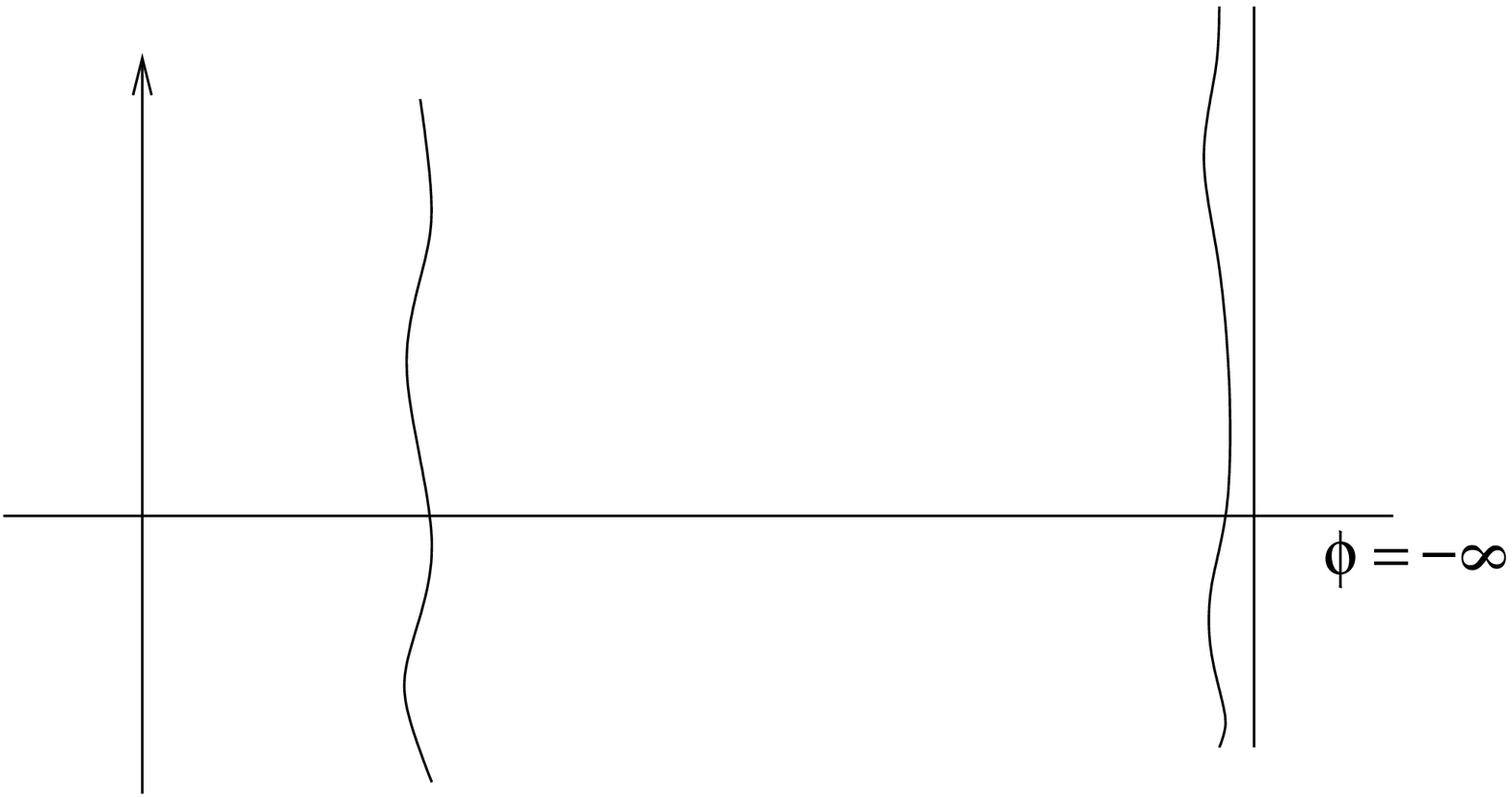}
\end{center}
\begin{center} Fig. 2 \end{center}

While linear response theory enables us to calculate transport
coefficients of fluid dynamics, we now briefly discuss non-linear
fluid dynamics and gravity, and indicate a remarkable connection
between the (relativistic) Navier-Stokes equations of fluid dynamics
and the long wavelength oscillations of the horizon of a black brane
which is described by Einstein's equations of general relativity with
a negative cosmological constant.

On general physical grounds a local quantum field theory at very high
density can be approximated by fluid dynamics.  In a conformal field
theory in $3+1$ dim. we expect the energy density $\epsilon \propto T^4$, where
$T$ is the local temperature of the fluid.  Hence fluid dynamics is a
good approximation for length scales $L \gg 1/T$.  The
dynamical variables of relativistic fluid dynamics are the four
velocities: $u_\mu (x)$ $(u_\mu u^\mu = -1)$, and
the densities of local conserved currents.  The conserved currents are
expressed as local functions of the velocities, charge densities and
their derivatives.  The equations of motion are given by the
conservation laws.  An example is the conserved energy-momentum tensor
of a charge neutral conformal fluid:
\be
T^{\mu\nu} = (\epsilon + P) u^\mu u^\nu + P\eta^{\mu\nu} -
\eta\left(P^{\mu\alpha} P^{\nu\beta}(\partial_\alpha u_\beta +
\partial_\beta u_\alpha) - {1\over3} P^{\mu\nu} \partial_\alpha
u^\alpha\right) + \cdots
\label{thirtythree}
\ee 
where $\epsilon$ is the energy density, $P$ the pressure, $\eta$
is the shear viscocity and $P^{\mu\nu} = u^\mu u^\nu + \eta^{\mu\nu}$.
These are functions of the local temperature.  Since the fluid
dynamics is conformally invariant (inheriting this property from the
parent field theory) we have $\eta_{\mu\nu} T^{\mu\nu} = 0$ which
implies $\epsilon = 3P$.  Since the speed of sound in the fluid is
given by $v^2_s = \d{\partial P \over \partial \epsilon}$, $v_s = \d{1
\over \sqrt{3}}$ or re-instating units $v_s = \d{c \over \sqrt{3}}$,
where $c$ is the speed of light in vacuum.  The pressure and the
viscosity are then determined in terms of temperature from the
microscopic theory.  In this case conformal symmetry and the
dimensionality of space-time tells us that $P \sim T^4$ and $\eta \sim
T^3$.  However the numerical coefficients need a microscopic
calculation. The Navier-Stokes equations are given by (\ref{thirtythree}) and
\be
\partial_\mu T^{\mu\nu} = 0
\label{thirtyfour}
\ee
The conformal field theory of interest to us is a gauge theory and a
gauge theory expressed in a fixed gauge or in terms of manifestly gauge
invariant variables is not a local theory.  In spite of this
(\ref{thirtythree}) seems to be a reasonable assumption and the local
derivative expansion in (\ref{thirtythree}) can be justified using the
AdS/CFT correspondence.  

We now briefly indicate that the eqns.(\ref{thirtythree}),
(\ref{thirtyfour}) can be deduced systematically from black brane
dynamics\cite{eighteen}.  Einstein's equation (\ref{twentyfive}) admits a boosted
black-brane solution 
\be
ds^2 = -2u_\mu dx^\mu dv - r^2 f(br)u_\mu u_\nu dx^\mu dx^\nu + r^2
P_{\mu\nu} dx^\mu dx^\nu
\label{thirtyfive}
\ee
where $v,r,x^\mu$ are in-going Eddington-Finkelstein coordinates and 
\bea
f(r) &=& 1 - {1 \over r^4} \nonumber \\ && \\
u^v &=& {1 \over \sqrt{1 - \beta^2_i}}, \ u^i = {\beta^i \over
\sqrt{1 - \beta^2_i}} \nonumber
\label{thirtysix}
\eea
where the temperature $T = 1/\pi b$ and the velocities
$\beta_i$ are all constants.  This 4-parameter solution can be
obtained from the solution with $\beta^i = 0$ and $b=1$ by  a boost
and a scale transformation.  The key idea is to make $b$ and $\beta^i$
slowly varying functions of the brane volume i.e. of the co-ordinates
$x^\mu$.  One can then develop a perturbative non-singular solution of
(\ref{twentyfive}) as an expansion in powers of $1/LT$.  Einstein's
equations are satisfied provided the velocities and pressure that
characterise (\ref{thirtyfive}) satisfy the Navier-Stokes
eqns. The pressure $P$ and viscosity $\eta$ can be
exactly calculated to be\cite{eighteen,nineteen}  
\be
P = (\pi T)^4 \ {\rm and} \ \eta = 2(\pi T)^3
\label{thirtyseven}
\ee
Using the thermodynamic relation $dP = sdT$ we get the entropy density
to be $s = 4\pi^4 T^3$ and hence obtain the famous equation of
Policastro, Son and Starinets, 
\be
{\eta \over s} = {1 \over 4\pi}
\label{thirtyeight}
\ee 
which is a relation between viscosity of the fluid and the entropy
density.  Strongly coupled fluid behaves more like a liquid than a
gas.  Systematic higher order corrections to (\ref{thirtythree})
can also be worked out.  

The experiments at RHIC seem to support very rapid thermalization and
a strongly coupled quark-gluon plasma with very low viscosity
coefficient, ${\eta \over s} \gsim {1 \over 4\pi}$.  

The fluid dynamics/gravity correspondence can also be used to study
non-equilibrium processes like thermelization which are dual to black
hole formation in the gravity theory.  An important result in this
study is that the thermalization time is more rapid than the expected
value $\propto {1 \over T}$ where $T$ is the temperature\cite{twenty}.

Another important result is the connection between the area theorems
of general relativity and the positivity of entropy in fluid
dynamics\cite{1one22}.

The fluid/gravity correspondence is firmly established for a $3+1$
dim. conformal fluid dynamics which is dual to gravity in AdS$_5$
space-time.  A similar connection holds for $2+1$ dim. fluids and
AdS$_4$ space-time.  We shall discuss the case of non-conformal fluid
dynamics in $1+1$ dim. separately.  A special (asymmetric) scaling
limit of the relativistic Navier-Stokes equations, where we send $v_s
= \d{c \over \sqrt{3}} \rightarrow \infty$ leads to the standard
non-relativistic Navier-Stokes equations for an incompressible
fluid\cite{twentyone}.  {\it In summary we have a truly remarkable
  relationship between two famous equations of physics viz. Einstein's
  equations of general relativity and the Navier-Stokes equations}.

Finally it is hoped that the AdS/CFT correspondence lends new
insights to the age old problem of turbulence in fluids.  Towards this goal
the AdS/CFT correspondence has also been established for forced
fluids, where the `stirring' term is provided by an external metric and
dilaton field\cite{twentytwo}.

\section{Non-conformal fluid dynamics in 1+1 dim. 
from gravity {\small \cite{twentythree,twentyfour}}}\label{aba:sec12}

The famous Policastro, Son and Starinets result (\ref{thirtyeight}) is
indeed a cornerstone of the gauge/gravity duality.  It was originally
derived in the context of conformal fluid dynamics.  However one
suspects that the conjectured bound ${\eta \over s} \geq {1 \over
  4\pi}$ may be more generally valid.  We present a summary of a
project of the fluid dynamics description, via the gauge/gravity
duality for the case of $N$ D1 branes at finite temperature $T$.  The
gauge theory describing the collective excitations of this system is a
$1+1$ dim. $SU(N)$ gauge theory with 16 supersymmetries.  Note that
this gauge theory is not conformally invariant.  At high temperatures
we expect the theory to have a fluid dynamics description, in terms of
a 2-velocity $u^\mu$ and stress tensor
\[
T^{\mu\nu} = (\epsilon + P) u^\mu u^\nu + P \eta^{\mu\nu} - \xi P^{\mu\nu}
\partial_\lambda u^\lambda
\]
Note that $\eta_{\mu\nu} T^{\mu\nu} = -3\xi \partial_\lambda
u^\lambda$, where $\xi$ is the bulk viscosity.  The dual gravity
description corresponds to 2 regimes.  For $\sqrt{\lambda} N^{-2/3}
\ll T \ll \sqrt{\lambda}$, the gravity dual is a classical solution
corresponding to a non-external D1 brane.  For $\sqrt{\lambda} N^{-1}
\ll T \ll \sqrt{\lambda} N^{-2/3}$ the gravity solution corresponds to
a fundamental string.  Here $\lambda = g^2_{YM} N$.

For both regimes we find the following exact answers for the strongly
coupled fluid dynamics.  There is exactly one gauge invariant
quasi-normal mode with dispersion:
\be
\omega = {q \over \sqrt{2}} - {i \over 8\pi T} q^2
\label{thirtynine}
\ee
The linearized fluid dynamics equations lead to the dispersion relation:
\be
\omega = v_s q - {i\xi \over 2(\epsilon + P)} q^2
\label{fourty}
\ee 
$v^2_s = {\partial P \over \partial \epsilon}$ is the velocity of
the sound mode.  Using $v^2_s = {1\over2}$ and the relation $\epsilon
+ P = Ts$, we once more arrive at 
\be
{\xi \over s} = {1 \over 4\pi}
\label{fourtyone}
\ee 
It is worth pointing out that (\ref{fourtyone}) is valid even if we work
with the geometry of D1 branes at cones over Sasaki-Einstein
manifolds.  Here the corresponding gauge theory is different from the
gauge theory with 16 supercharges that we mentioned before.  Using
similar techniques we have also studied the case of the $SU(N)$ gauge
theory in $1+1$ dim. with finite $R$-charge density.  The dual
supergravity solution is that of a non-extremal D1 brane spinning
along one of the Cartan directions of $SO(8)$ which reflects the isometry of
$S^7$ present in the near horizon geometry.  In this case, besides
energy transport, there is also charge transport.  The transport
coefficients like electrical and heat conductivity can be
calculated, and the Weidemann-Franz law can be verified.  Once again (14) 
is valid. 
\bigskip

\section{A New Term in Fluid dynamics:}

The fluid dynamics of a charged fluid is described by the conserved
stress tensor $T_{\mu\nu}$ and charged current $J_\mu$.  The
constituent equations are (to leading order in the derivative
expansion)
\be
T_{\mu\nu} = P(\eta_{\mu\nu} + 4u_\mu u_\nu) - 2\eta \sigma_{\mu\nu} + \cdots 
\label{fourtytwo}
\ee
\be
J_\mu = n u_\mu - D P_\mu^\nu D_\nu n.
\label{fourtythree}
\ee 
where $n$ is the charge density.  However in the study of the
charged black brane dual to a fluid at temperature $T$ and chemical
potential $\mu$, a new term was discovered in the charged current \be
J_\mu = n u_\mu - D P_\mu^\nu D_\nu n + \zeta \ell_\mu
\label{fourtyfour}
\ee 
$\ell_\mu = \epsilon_{\mu\nu\rho\sigma} u_\nu
\omega_{\rho\sigma}$, $\omega_{\rho\sigma} = \partial_\rho u_\sigma -
\partial_\sigma u_\rho$ (the vorticity).  The appearance of the new
voriticity induced current in (\ref{fourtyfour}) is directly related
to the presence of the Chern-Simons term in the Einstein-Maxwell
lagrangian in the dual gravity description\cite{twentyfive,twentysix}.

In a remarkable paper Son and Surowka\cite{twentyseven} showed that
the vorticity dependent term in (\ref{fourtyfour}) always arises in a
relativistic fluid dynamics in which there is an anomalous axial
$U(1)$ current: $\partial_\mu J_\mu^A = -{1 \over 8}
CF_{\mu\nu} F_{\rho\sigma} \epsilon_{\mu\nu\rho\sigma}$.  They showed
on general thermodynamic grounds that
\[
\zeta = C\left(\mu^2 - {2\over3} {\mu^3 n \over \epsilon + P}\right)
\]
where $\epsilon$ and $P$ are the energy density and pressure, and
$\mu$ is the chemical potential. If $C = 0$ one recovers the result
(\ref{fourtythree}) of Landau and Lifshitz.  This new term may be
relevant in understanding bubbles of strong parity violation observed
at RHIC and generally in the description of rotating charged fluids.

\section{Implications of the gauge fluid/gravity correspondence for the 
information paradox of black hole physics}

The Navier-Stokes equations imply dissipation and violate time
reversal invariance.  The scale of this violation is set by
$\eta/\rho$ ($\eta$ is the viscosity and $\rho$ is the density) which
has the dim. of length (in units where the speed of light $c = 1$).
There is no paradox here with the fact that the underlying theory is
non-dissipative and time reversal invariant, because we know that the
Navier-Stokes equations are not a valid description of the system for
length scales $\ll \eta/\rho$, where the micro-states should be taken
into account.  {\it An immediate important implication of this fact
  via the AdS/CFT correspondence is that there will always be
  information loss in a semi-classical treatment of black holes in
  general relativity.}  This fact raises an important question: while
we understand that information loss in fluid dynamics because we know
the underlying constitutent gauge theory, a similar level of
understanding does not exist on the string/gravity side, because we as
yet do not know the exact equations for all values of the string
coupling.

\section{Concluding remarks:}

In this note we have reviewed the emergence, via the AdS/CFT
correspondence, of a quantum theory of gravity from an interacting
theory of D-branes. Besides giving a precise definition of quantum
gravity in terms of non-abelian gauge theory, this correspondence
turns out to be a very useful tool to calculate properties of strongly
coupled gauge theories using semi-classical gravity. The
correspondence of dynamical horizons and the fluid dynamics limit of the
gauge theory enables calculation of transport coefficients like
viscosity and conductivity. We also indicated
that dissipation in fluid dynamics implies that in semi-classical
gravity there will always be `information loss'.

We conclude with a brief mention of other applications of the AdS/CFT
correspondence to various problems in physics.
\medskip

\noindent {\bf Condensed Matter:}
\medskip

The AdS/CFT correspondence offers a tool to explore many questions
in strongly coupled condensed matter systems in the vicinity of a
quantum critical point. It enables calculation of transport properties,
non-fermi liquid behavior, quantum oscillations and properties of 
fermi-surfaces etc.\cite{twentyeight,twentynine,thirty,thirtyone}. There is a
puzzling aspect in the application of semi-classical gravity to
condensed matter systems:  what determines the smallness of the
gravitational coupling, which in the gauge theory goes as $N^{-2}$?

Another interesting development is bulk superconductivity i.e. the
presence of a charged scalar condensate in a black hole
geometry\cite{thirtytwo}. This has interesting implications for
superfluidity in the quantum field theory on the boundary.
\medskip

\noindent {\bf QCD and Gauge theories:}
\medskip

The AdS/CFT correspondence is a powerful tool to calculate multi-gluon
scattering amplitudes in ${\cal N} = 4$ gauge theories in 3+1 dims. This
is done by relating the amplitude to the calculation of polygonal Wilson
lines in a momentum space version of $AdS_5$.\cite{thirtythree,thirtyfour}

Even though the basic theory of the quark-gluon plasma is QCD,
calculations in ${\cal N} = 4$ gauge
theories do indicate qualitative agreement with RHIC observations. We
have already remarked that the observed value of ${\eta \over s}$ in
(11) is in qualitative agreement with RHIC data. Another
calculation of interest is that of jet quenching which corresponds
to computing the drag force exerted by a trailing string attached to a
quark on the boundary, in the presence of a AdS black hole\cite{thirtyfive}.

The AdS/CFT correspondence has also yielded a geometric understanding
of the phenomenon of chiral symmetry breaking\cite{2two34,3three35}.
\medskip

\noindent {\bf Singularities in Quantum Gravity}
\medskip

The AdS/CFT correspondence provides a way to discuss the quantum
resolution of the singularities of classical general relativity.  One
strategy would be to study the resolution of singularities that occur
in the gauge theory in the $N \rightarrow \infty$ limit.  Among these
are singularities corresponding to transitions of order greater than
two\cite{4four37,5five38} which admit a resolution in a double scaling
limit.  The difficult part here is the construction of the map between
the gauge theory simgularity and the gravitational singularity.  A
proposal in the context of the Horowitz-Polchinski cross-over was made
in reference 42.

\section{Acknowledgement:}

I would like to thank K.K. Phua and Belal Baaquie for their warm
hospitality and Gautam Mandal for a critical reading of the draft and 
very useful discussions.

\end{document}